\documentclass{elsart}

\newcommand{\AX}{\ensuremath{\tilde{A}^1A_2 \leftarrow \tilde{X}^1A_1} }

\makeatletter % <-- bitte in FAQ nachsehen
\newcommand{\roem}[1]{
        \textsc{% für Kapitälchen; hier z.B. auch \textrm oder so
                \@roman{#1.}% Das Argument als roemische Ziffer
        }\ % Ein normalgroßes Leerzeichen, wichtig bei nonfrenchspacing
        \ignorespaces % Keine weiteren Leerzeichen ausgeben.
}
\newcommand{\Roem}[1]{
        \textsc{% für Kapitälchen; hier z.B. auch \textrm oder so
                \@Roman{#1.}% Das Argument als roemische Ziffer
        }\ % Ein normalgroßes Leerzeichen, wichtig bei nonfrenchspacing
        \ignorespaces % Keine weiteren Leerzeichen ausgeben.
}
\makeatother % <-- bitte in FAQ nachsehen

\showhyphens{rovibrational}
\showhyphens{formaldehyde}

\usepackage{graphicx} % Einbinden von Grafiken

\usepackage{subfigure}

\usepackage[latin1]{inputenc}

\begin{document}

\begin{frontmatter}



\title{Spectroscopy of the $\tilde{A}^1A_2\leftarrow \tilde{X}^1A_1$ transition of formaldehyde in the 30140--30790\,cm$^{-1}$ range: the $2^1_0 4^3_0$ and $2^2_0 4^1_0$ rovibrational bands}


\author[mpq]{Michael Motsch},
\author[mpq]{Markus Schenk},
\author[mpq]{Martin Zeppenfeld},
\author[hhdd]{Michael Schmitt},
\author[nij]{W. Leo Meerts},
\author[mpq]{Pepijn W.H. Pinkse\corauthref{cor1}},
%\ead{pepijn.pinkse@mpq.mpg.de}
%\ead[url]{http://www.mpq.mpg.de/qdynamics}
\author[mpq]{Gerhard Rempe}
\corauth[cor1]{pepijn.pinkse@mpq.mpg.de}
\address[mpq]{Max-Planck-Institut f{\"u}r Quantenoptik, Hans-Kopfermann-Str. 1, 85748 Garching, Germany}
\address[hhdd]{Heinrich-Heine Universit{\"a}t, Institut f\"ur Physikalische Chemie, Universit{\"a}tsstr. 1, 40225 D{\"u}sseldorf, Germany}
\address[nij]{Molecular- and Biophysics Group, Institute for Molecules and Materials, Radboud University Nijmegen, P.O. Box 9010, 6500 GL Nijmegen, The Netherlands}

\begin{abstract}
Room-temperature absorption spectroscopy of the $\tilde{A}^1A_2\leftarrow \tilde{X}^1A_1$ transition of formaldehyde has been performed in the 30140--30790\,cm$^{-1}$ range allowing the identification of individual lines of the $2^1_0 4^3_0$ and $2^2_0 4^1_0$ rovibrational bands. Using tunable ultraviolet continuous-wave laser light, individual rotational lines are well resolved in the Doppler-broadened spectrum.
Making use of genetic algorithms, the main features of the spectrum are reproduced.
Spectral data is made available as Supporting Information.
\end{abstract}

\end{frontmatter}

%%%%%%%%%%%%%%%%%%%
%%% The authors %%%
%%%%%%%%%%%%%%%%%%%
%\author{Michael Motsch}
%\affiliation{Max-Planck-Institut f{\"u}r Quantenoptik, Hans-Kopfermann-Str. 1, 85748 Garching, Germany}
%\author{Markus Schenk}
%\affiliation{Max-Planck-Institut f{\"u}r Quantenoptik, Hans-Kopfermann-Str. 1, 85748 Garching, Germany}
%\author{Martin Zeppenfeld}
%\affiliation{Max-Planck-Institut f{\"u}r Quantenoptik, Hans-Kopfermann-Str. 1, 85748 Garching, Germany}
%\author{Michael Schmitt}
%\affiliation{Heinrich-Heine Universit{\"a}t, Institut f\"ur Physikalische Chemie, Universit{\"a}tsstr. 1, 40225 D{\"u}sseldorf, Germany}
%\author{W. Leo Meerts}
%\affiliation{Molecular- and Biophysics Group, Institute for Molecules and Materials, Radboud University Nijmegen, P.O. Box 9010, 6500 GL Nijmegen, The Netherlands}
%\author{Pepijn W.H. Pinkse}
%\email{pepijn.pinkse@mpq.mpg.de}
%\affiliation{Max-Planck-Institut f{\"u}r Quantenoptik, Hans-Kopfermann-Str. 1, 85748 Garching, Germany}
%\author{Gerhard Rempe}
%\affiliation{Max-Planck-Institut f{\"u}r Quantenoptik, Hans-Kopfermann-Str. 1, 85748 Garching, Germany}

%\title{Spectroscopy of the $\tilde{A}^1A_2\leftarrow \tilde{X}^1A_1$ transition of formaldehyde in the 30140--30800\,cm$^{-1}$ range: the $2^1_0 4^3_0$ and $2^2_0 4^1_0$ rovibrational bands}

%\date{\today, \thistime, PREPRINT}

%%%%%%%%%%%%%%%%
%%% Abstract %%%
%%%%%%%%%%%%%%%%
%\begin{abstract}
%\end{abstract}
%
%\maketitle

% main text

%%%%%%%%%%%%%%%%%%%%
%%% Introduction %%%
%%%%%%%%%%%%%%%%%%%%
\section{Introduction}

The ultraviolet (UV) spectrum of formaldehyde (H$_2$CO) has been extensively studied already around the birth of molecular spectroscopy \cite{Dieke1934}. As one of the simplest polyatomic molecules it can be considered a model system for molecular physics.

A variety of studies has been carried out to experimentally determine key parameters for formaldehyde photochemistry such as photodissociation quantum yields and absolute absorption cross sections. These numbers are relevant for atmospheric science, for which formaldehyde is an important molecule. Formaldehyde is present in the atmosphere at concentration of $\sim$\,50\,pptv (parts per trillion by volume) in clean tropospheric air \cite{Riedel1999} and up to 10--70\,ppbv (parts per billion by volume) in the air in urban centers \cite{Williams1996,Grosjean1983}. By excitation of the $\tilde{A}^1A_2\leftarrow \tilde{X}^1A_1$ transition in the 260--360\,nm wavelength range two dissociation channels \cite{Moore1983,Bowman2006} with high quantum yields \cite{McQuigg1969,Clark1978,Horowitz1978,Horowitz1978a,Reilly1978,Moortgart1978} are open: (a)
$\mathrm{H_2CO} + h\nu \rightarrow \mathrm{H_2 + CO}$ and (b) $\mathrm{H_2CO} + h\nu \rightarrow \mathrm{H + HCO}$. The reaction channel (a) opens at wavelengths $<$360\,nm, reaction channel (b) opens at wavelengths $<$330\,nm.

Formaldehyde is also an interesting molecule for the field of cold polar molecules \cite{Doyle2004,Bethlem2003}. Chemical reactions at low collision energies become controllable by externally applied electric fields \cite{Avdeenkov2003,Krems2005,Avdeenkov2006}. An example is the hydrogen abstraction channel in the reaction of formaldehyde with hydroxyl radicals \cite{Hudson2006}. Due to its large dipole moment and large linear Stark shift \cite{Townes:Microwave}, cold formaldehyde molecules can be prepared by electrostatic filtering from a thermal gas where fluxes of up to $10^{10}\,\mathrm{s^{-1}}$ with a mean velocity of $50\,\mathrm{m/s}$ corresponding to a translational temperature of $\approx$\,5\,K have been demonstrated \cite{Rangwala2003,Junglen2003}. The rich ultraviolet spectrum in the range 260--360\,nm \cite{Herzberg:3,Brand1956,Moule1975,Clouthier1983,Cantrell1990,Pope2005,Smith2006} gives the opportunity to spectroscopically study these velocity-filtered cold guided molecules with high resolution, since this wavelength region is accessible with narrow-bandwidth frequency-doubled continuous-wave (cw) dye lasers. To address individual rotational transitions of these velocity-filtered molecules, it is necessary to predict line positions for states populated in the guided beam with high accuracy, requiring refined rotational constants. The work described in this paper is a necessary prologue to experimentally access the internal state distribution of the slow guided molecules \cite{Motsch2007}.

In this paper we give a detailed description of the experimental setup used for measurements of (weak) absorption spectra in the near-UV spectral region. We compare our measurements to previous data and show the improvement in resolution, which allows identification and precise determination of line positions for individual rotational transitions. The lineshape of isolated lines is well reproduced by a room-temperature Doppler profile. We find deviations to line positions calculated with literature values for rotational constants \cite{Clouthier1983,Smith2006}. Using genetic algorithms \cite{Meerts2004,Meerts2006a,Meerts2006} for fitting rotational constants of the $2^1_0 4^3_0$ and $2^2_0 4^1_0$ rovibrational bands, good agreement between the simulation and the measured spectra is achieved over a wide range of the spectrum. However, the region between 30390--30410\,cm$^{-1}$ had to be excluded from the simulation, which might indicate the presence of perturbations. We briefly review how the used setup can be modified for Doppler-free measurements on selected lines of this weak transition \cite{Zeppenfeld2007}. Doppler-free linewidths of 40--50\,MHz give a lower bound for the lifetime of the energy levels, which are known to predissociate.

%%%%%%%%%%%%%%%%%%%%%%%%%%
%%% Experimental Setup %%%
%%%%%%%%%%%%%%%%%%%%%%%%%%
\section{Experimental Setup}
\begin{figure}
\centering
\includegraphics[width=0.95\columnwidth]{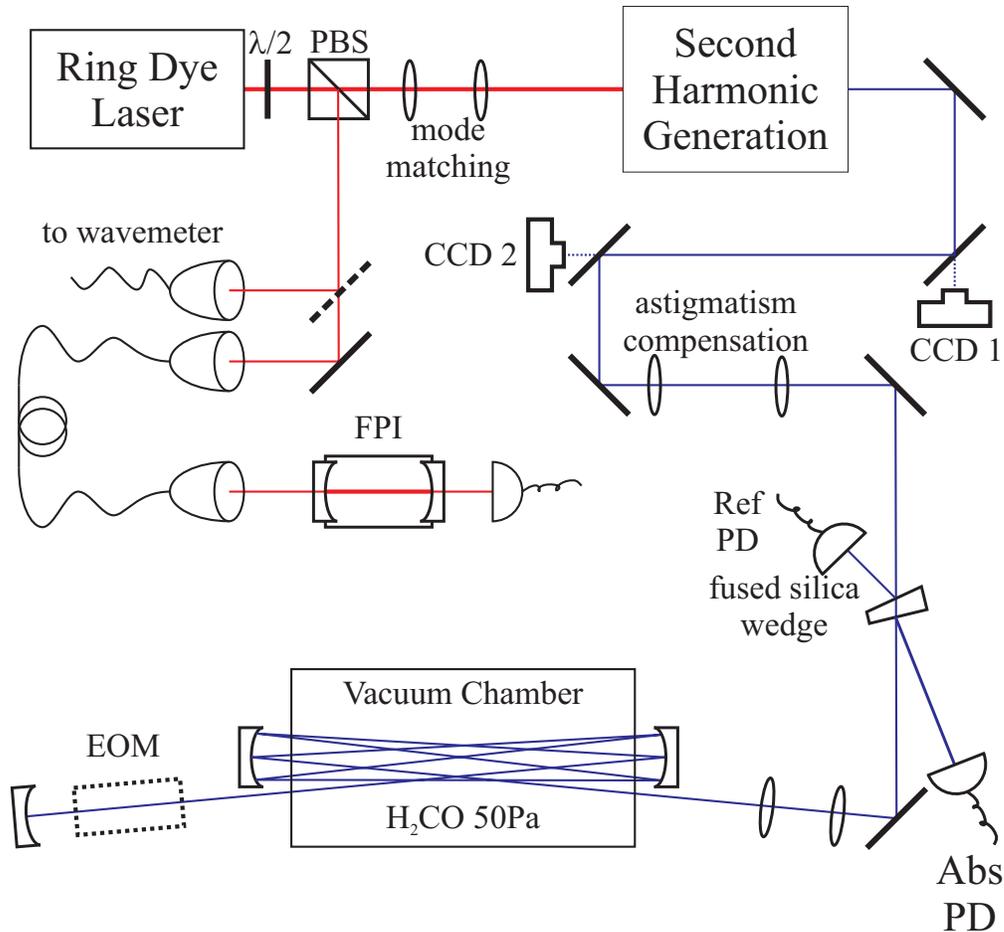}
\caption{The experimental setup used for room-temperature absorption spectroscopy including laser system and vacuum chamber. A part of the fundamental laser light from the dye laser is sent through optical fibres to a wavemeter and a Fabry-Perot-Interferometer (FPI) for wavelength measurement and calibration of the scan speed. UV laser light is generated using a BBO crystal placed in an external enhancement cavity. After passing the beam pointing correction system, consisting of two CCD cameras and two steering mirrors, and the astigmatism compensation setup, the UV light is sent through the formaldehyde spectroscopy chamber using a multipass setup with an overall path length of 3.15\,m. Reference and absorption signals are measured using the reflection from a fused-silica wedge placed near normal incidence. For frequency-modulated Doppler-free measurements an electro-optical modulator (EOM, indicated by the dashed box) is inserted into the retroreflected beam for frequency modulation.}
\label{pic:ExpSetup}
\end{figure}

The experimental setup used for our absorption spectroscopy measurements is standard \cite{Demtroeder:Laserspectropscopy}. However, due to the small absorption cross sections of formaldehyde ($\sigma\sim10^{-19}$\,cm$^2$), measurements must be performed at relatively high densities and long optical path lengths. As shown in Fig.\,\ref{pic:ExpSetup}, a home-made multipass setup using 2 spherical UV mirrors placed outside the vacuum chamber and one additional retro reflection mirror was used.

\subsection{Laser System}
Tunable narrow-bandwidth ultraviolet laser light is produced by second-harmonic generation of the output from a continuous-wave ring dye laser (Coherent 899, pumped by a Coherent VERDI) in an external enhancement cavity (Coherent MBD-200). The dye laser is operated with the dye "DCM special" using a stainless steel nozzle optimized for high pressures (9\,bar). The frequency of the dye laser is locked to an external temperature-stabilized reference cavity resulting in a linewidth of $\approx$\,0.5\,MHz. A part of the fundamental laser light is split off and sent via optical fibres to a wavemeter (HighFinesse WS7) and to an Invar Fabry-Perot-Interferometer (FPI) with a free spectral range of 1.0024\,GHz. The wavemeter used in our experiments has a specified absolute frequency uncertainty of 100\,MHz when properly calibrated to a light source of well known frequency, in our case a stabilized HeNe laser (SIOS SL03). We therefore carefully estimate the absolute frequency accuracy of our measurements to be better than 300\,MHz = 0.01\,cm$^{-1}$ in the fundamental and hence 0.02\,cm$^{-1}$ in the UV. At wavelengths around 330\,nm typical UV output powers are $\approx$\,20--30\,mW at fundamental powers of 300--400\,mW. Measurements were performed with UV power around 20\,mW.

The generated UV light is sent through two cylindrical lenses for astigmatism compensation, which is intrinsic to type II critical phase matching in BBO crystals used for second-harmonic generation at this wavelength. Furthermore, the residual transmission of UV light through two 45\,$^\circ$ steering mirrors is monitored on two CCD cameras, which allows compensation of beam pointing effects when changing the laser wavelength. Without realignment, changes of the fundamental wavelength by 0.5\,cm$^{-1}$ lead to a significant difference in signal on the absorption and reference photodiodes due to the large optical path length. By manual realigning of the laser beam onto the two CCD cameras after each wavelength change, no more beam pointing effects were observed.

\subsection{Formaldehyde Spectroscopy Setup}
Absorption spectroscopy is performed in a vacuum chamber of $\sim$22.5\,cm length. % 22.5cm is given as size in Zeppenfeld et al. 220mm Octagon size, the viewports have probably each 5mm length
Connected to the vacuum chamber are a turbo molecular pump, a pressure gauge (Pfeiffer Vacuum Compact Full Range Gauge PKR261), a flow valve for formaldehyde input and a flow valve allowing analysis of the chambers contents by a mass spectrometer (Pfeiffer Vacuum Prisma QMS200). The effective pumping speed of the turbo molecular pump can be reduced by a angle valve placed between the recipient and the pump to avoid excess pumping of formaldehyde. With the turbo molecular pump the spectroscopy chamber could be evacuated to a base pressure in the 10$^{-6}$\,mbar range.

Formaldehyde is produced by heating Paraformaldehyde (Sigma-Aldrich) to a temperature of 80--90\,$^\circ$C. To clean the dissociation products and remove unwanted water and polymer rests, the gas is led through a dry-ice cold trap at a temperature of $\approx$\,-80$^\circ$C \cite{Spence1935}. Without the cold trap the viewports of the vacuum chamber became coated with a white layer of paraformaldehyde after several hours of operation. With the cold trap this effect was no longer observable even after extensive use. Since formaldehyde molecules dissociate upon UV excitation, a stable flow of formaldehyde was maintained by slightly opening the valve between the turbo pump and the vacuum chamber. The partial pressures of formaldehyde and its dissociation products were monitored with the mass spectrometer and the formaldehyde input flow rate, as well as the flow to the turbo molecular pump optimized for a constant ratio. In this way measurements could be performed at a constant formaldehyde concentration. Measurements were performed at a constant pressure of 50\,Pa in the vacuum chamber with a formaldehyde fraction estimated to be $\geq$50\,\%.

To achieve a large optical path length in a relatively compact setup, a multipass and retro-reflection configuration was used. The UV laser beam is initially focussed into the vacuum chamber with the beam parameters mode-matched to the effective cavity mode generated by the two multipass mirrors. Two curved mirrors with a radius of curvature (RoC) of 200\,mm outside the vacuum chamber are used to refocus the beam into the vacuum chamber, allowing 7 passes. Using an additional mirror (RoC\,=\,500\,mm) for retro-reflection, the effective path length can even be doubled, yielding $\mathrm{\sim3.15\,m}$ in a compact $\mathrm{\sim22.5\,cm}$ length vacuum chamber. All curved mirrors are positioned such that their radius of curvature matches the Gaussian mode leaving the spectroscopy setup.

For detection a fused-silica wedge was placed near normal incidence which allows picking up part of the original beam used as power reference and part of the retro-reflected beam containing the absorption signal. The picked-up beams are focussed on UV sensitive Si photodiodes (Thorlabs PDA36EC).% with a size of 0.8$\times$0.8\,mm$^2$ (Thorlabs {\color{red}PDA10???}).%We prefer this method to the combination of a polarizing beamsplitter and a quarter-wave-plate due to the large wavelength tuning range of the dye laser and to inexplicable fluctuations in output polarization.

A linear voltage ramp was applied to the external scan input of the dye laser for sweeping the laser frequency over 20\,GHz in the fundamental, resulting in a scan speed of $\approx$\,45\,GHz/s. For each of these sweeps the central frequency of the fundamental was measured with the wavemeter before and after the sweep, giving agreement within 0.001\,cm$^{-1}$ in the fundamental. %{\color{red}(Or 0.0002\,cm$^{-1}$, if a better sweep cycle with less disturbance of the dye laser locking was used, but that was not the case for these measurements in the first vibrational band, we found out how we can do better later.)}
The FPI transmission was monitored for calibration of the scan speed. Subsequent scans were performed with a central frequency difference of 10\,GHz in the fundamental, giving $\approx$\,50\,\% overlap for concatenating individual sweeps.

\subsection{Data Acquisition and Data Analysis}
For data acquisition a 4-channel digital oscilloscope was used. Simultaneously, the external ramp, the FPI transmission peaks, the reference and absorption photodiode signals were recorded. For later data analysis the channels were rebinned to a resolution of 100\,MHz in the UV, which was chosen such that individual rotational transitions, which are Doppler-broadened to 2.4\,GHz, could be well resolved. Overlapping adjacent scans using the central frequencies measured with the wavemeter showed good overlap between lines present in both scans and confirmed the central wavelength measurements with the wavemeter.

\subsection{Modifications for Doppler-free measurements}
The experimental setup used for Doppler-free measurements and the developed analytical model for the amplitude of Doppler-free peaks is described in detail elsewhere \cite{Zeppenfeld2007}. Here only the modifications to the spectroscopy setup are summarized. To ease the detection of weak Doppler-free signals, frequency modulation (FM) spectroscopy \cite{Bjorklund1980,Bjorklund1983} is performed. For this an electro-optical modulator (EOM, Leysop EM400K) resonantly driven with a frequency of 15.8\,MHz for the creation of sidebands is placed in the retroreflected beam (see Fig. \ref{pic:ExpSetup}). Since a demodulation of the signal at 15.8\,MHz is necessary, fast photodiodes with a sufficiently high bandwidth are used for detection (Thorlabs PDA155). Furthermore, since the \AX electronic transitions in formaldehyde are weak, high laser powers are needed to reach a significant saturation and discriminate the Lamb dips against the Doppler-broadened background. After careful tuning of the laser system, UV laser powers of 250--350\,mW were available for this saturation-spectroscopy experiment.

%%%%%%%%%%%%%%%
%%% Results %%%
%%%%%%%%%%%%%%%
\section{Results and Discussion}
Previous measurements of the \AX rovibrational band of formaldehyde aimed at determination of absolute temperature-dependent absorption cross sections \cite{Cantrell1990,Pope2005,Smith2006}. Other experiments studied the quantum yield of the dissociation processes following the UV excitation \cite{McQuigg1969,Clark1978,Horowitz1978,Horowitz1978a,Reilly1978,Moortgart1978}. These cross sections and quantum yields are important parameters for the description of photochemistry in the atmosphere induced by the sunlight as discussed in the introduction. These measurements have been performed using either broadband light sources and spectrometers (\cite{Cantrell1990} and refs. therein) with resolutions above 1\,cm$^{-1}$ or pulsed lasers with spectral resolutions of, e.g., 0.35\,cm$^{-1}$ \cite{Pope2005,Smith2006}. An exception to this are the measurements by Co \emph{et al.} \cite{Co2005} with a resolution of 0.027\,cm$^{-1}$, spanning the long wavelength range (351--356\,nm) of the \AX band, and the measurements of Schulz \emph{et al.} of the $2^1_04^1_0$ and $2^2_04^1_0$ rovibrational band \cite{SchulzThesis}. Except for these two experiments the resolution of previous measurements is not high enough to resolve individual rotational lines with a Doppler width of 2.4\,GHz at room temperature.

\subsection{Comparison to previous studies}
\begin{figure}
\centering
%%% Origin
\includegraphics[width=0.95\columnwidth]{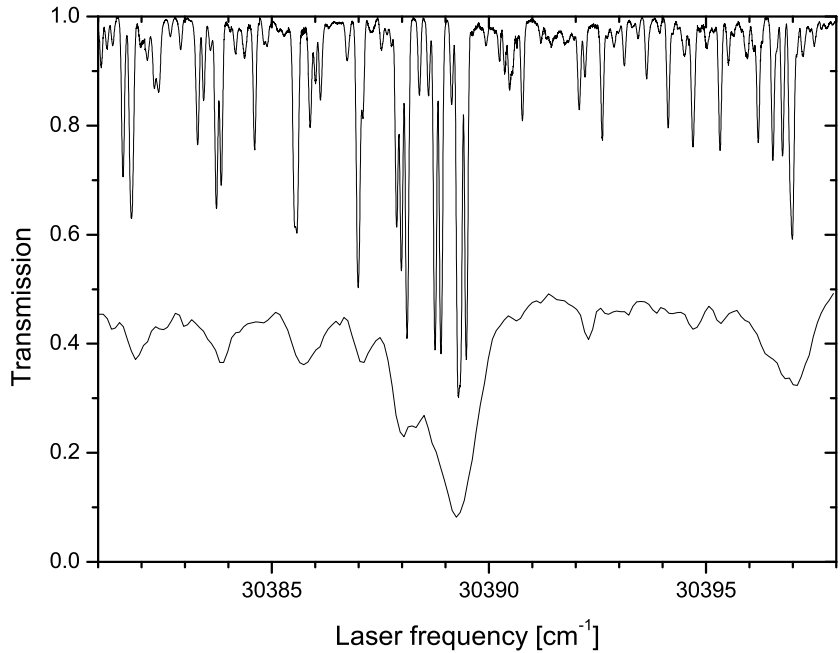}
\caption{Comparison in resolution of our absorption data (top) to previous measurements (bottom). The measurements by Smith \emph{et al.} \cite{Smith2006} with the so-far reported highest resolution in this wavelength range are chosen as reference. A wavelength range with many lines close together, converging to the bandheads of the $^rR\,K^{''}_{a}$\,=\,3 progression at $\approx$\,30389\,cm$^{-1}$ and $^rR\,K^{''}_{a}$\,=\,4 progression at $\approx$\,30397\,cm$^{-1}$, was selected to show how individual lines can be resolved now. The data of Smith \emph{et al.} are offset by 0.5 for clarity.}
\label{pic:SpecHiRes}
\end{figure}

The improvement in resolution compared to previous studies \cite{Smith2006} with a resolution of 0.35\,cm$^{-1}$ is shown in Fig. \ref{pic:SpecHiRes}, where the region around the band heads of the $^rR\,K^{''}_{a}$\,=\,3 progression at $\approx$\,30389\,cm$^{-1}$ and the $^rR\,K^{''}_{a}$\,=\,4 progression at $\approx$\,30397\,cm$^{-1}$ with many lines close together are shown. Individual rotational lines are well resolved and accurate line positions as well as intensities can be determined. For the figures shown (also for Supplementary Information) our measured data was binned to a resolution of 100\,MHz = 0.003\,cm$^{-1}$ which is sufficient to resolve the Doppler-broadened lines.%The wavemeter used in our experiments to measure the dye laser fundamental frequency has a specified absolute frequency uncertainty of 100\,MHz when properly calibrated to a light source of well known frequency, in our case a stabilized HeNe laser. {\color{red}We therefore carefully estimate the absolute frequency accuracy of our measurements to be better than 300\,MHz = 0.01\,cm$^{-1}$ in the fundamental and hence 0.014\,cm$^{-1}$ in the UV.}

Smith et al. \cite{Smith2006} used a lineshape function with a full width at half maximum (FWHM) of 0.45\,cm$^{-1}$ to reproduce their measured spectra from simulations. This was surprising since the UV linewidth of their laser sources was expected to be $\leq$\,0.20\,cm$^{-1}$ as calculated from the measured linewidths of the fundamental. It was speculated that this could either be explained by extremely short excited state lifetimes not in agreement with literature values \cite{Moore1983} or by an additional technical broadening. In our measurements we find for isolated lines across the whole studied range a width of 2.4\,GHz FWHM, which is in good agreement with a Doppler-broadened line profile at room temperature (293\,K). This therefore rules out these extremely short lifetimes and confirmes their assumption of a larger laser linewidth.

\subsection{Fit of the $2^1_04^3_0$ and $2^2_04^1_0$ rovibrational band}
\begin{figure}
\centering
%\subfigure {\includegraphics[width=0.49\columnwidth]{CompSim30320.eps}}\hfill
%\subfigure {\includegraphics[width=0.49\columnwidth]{CompSim30325.eps}}
%\subfigure {\includegraphics[width=0.49\columnwidth]{CompSim30330.eps}}\hfill
%\subfigure {\includegraphics[width=0.49\columnwidth]{CompSim30335.eps}}
%\subfigure {\includegraphics[width=0.49\columnwidth]{CompSim30340.eps}}\hfill
%\subfigure {\includegraphics[width=0.49\columnwidth]{CompSim30345.eps}}
%\subfigure {\includegraphics[width=0.49\columnwidth]{CompSim30350.eps}}\hfill
%\subfigure {\includegraphics[width=0.49\columnwidth]{CompSim30355.eps}}
\includegraphics[width=\textwidth]{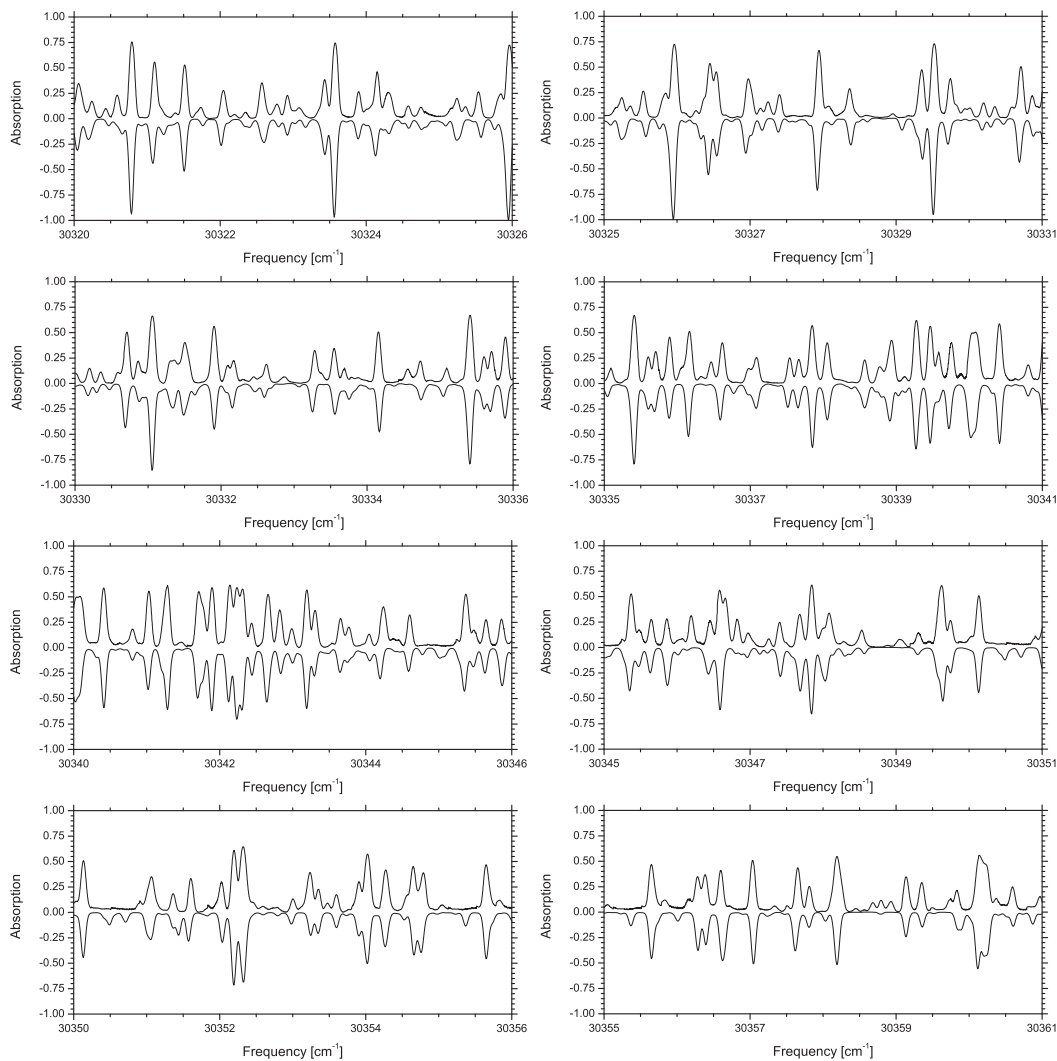}
\caption{Comparison between measured absorption (top) and simulation (inverted, bottom) around the origin of the $2^1_04^3_0$ vibrational band at 30340.08\,cm$^{-1}$.}
\label{pic:CompSim}
\end{figure}

For the fit of the molecular parameters to the experimental spectrum, Watson's A-reduced Hamiltonian \cite{Watson1967,Watson1968a}, including quartic and sextic centrifugal distortion terms has been used with a genetic algorithm (GA) as optimizer.  Details about the GA and the cost function used for evaluation of the quality of the fit can be found in Refs. \cite{Meerts2004,Meerts2006a,Meerts2006}. Table \ref{tab:fit} compiles the so-determined parameters and compares them to the previous parameters for the ground state \cite{Clouthier1983} and the excited vibronic states \cite{Smith2006}. The ground-state parameters are in excellent agreement with the values that have been deduced from microwave frequencies and combination differences in infrared and electronic spectra, weighted by appropriate factors \cite{Clouthier1983}. The parameters for the $2_0^1 4_0^3$ and $2_0^2 4_0^1$ vibronic bands are compared to the values given by Smith \textit{et al.}.  To improve the quality of the fit, the experimental data above 30390\,cm$^{-1}$ have been excluded from the fit of the $2_0^1 4_0^3$ vibronic band, since strong perturbations of the line positions between 30390 and 30404\,cm$^{-1}$ were observed. The reason why exclusion of a part of the data was necessary might be an interference with the weak vibronic $2^1_04^1_06^1_0$ band with its origin at 30395\,cm$^{-1}$. Although we included this band with the molecular parameters from \cite{Apel1985} as starting values in our fit, no improvement of the cost function could be obtained by this combined fit. Due to the high temperature, high $J$-states are populated and sextic centrifugal distortion terms in the fit have shown to be necessary. The appropriate nuclear spin statistics ($K_{a}$ even levels = 1; $K_{a}$ odd levels = 3) have been taken into account. For comparison, parts of the simulated and the measured absorption spectrum around the band origin are shown in Fig. \ref{pic:CompSim}. The simulated and measured spectrum as well as a complete line list of the calculated transitions is given in the supplementary material.

\clearpage

\begin{table*}[hbp]
\caption{Molecular parameters of the ground state and the $\tilde{A}^1$ excited state of formaldehyde
from a GA-Fit of the $2_0^1 4_0^3$ and $2_0^2 4_0^1$ vibronic bands. All values are given in MHz.}
\label{tab:fit}
{
\scriptsize
\begin{center}
\begin{tabular}{lrrccc}
\hline\hline
                    & GA-Fit $2_0^1 4_0^3$ & GA-Fit $2_0^2 4_0^1$      & Ref. \cite{Clouthier1983} & $2_0^14_0^3$Ref.\cite{Smith2006} & $2_0^24_0^1$Ref.\cite{Smith2006}\\
\hline
$A''$               &  281970.85(27)    &  281971.44(43)    & 281970.572(24)    &-&-\\
$B''$               &   38836.53(27)    &   38836.14(25)    & 38836.0455(13)    &-&-\\
$C''$               &   34002.67(16)    &   34002.57(25)    & 34002.2034(12)    &-&-\\
$10^3  D_{J}''$     &       75.38(7)    &    75.61(11)      &     75.295(21)    &-&-\\
$10^3  D_{JK}''$    &    1290.24(20)    &    1290.00(29)    &    1290.50(37)    &-&-\\
$10^3  D_{K}''$     &  19424.6(6)       & 19424.1(6)        &       19423(7)    &-&-\\
$10^3  d_{J}''$     &       9.85(24)    &      10.30(25)    &     10.4567(9)    &-&-\\
$10^3  d_{K}''$     &    1028.39(60)    &    1027.68(66)    &    1026.03(25)    &-&-\\
$10^9  H_{J}''$     &       33.5(16)    & 34.1(20)          &         31(21)    &-&-\\
$10^9   H_{JK}''$   & 29021(1)          & 29022(1)          &     29019(690)    &-&-\\
$10^9   H_{KJ}''$   & -112184(3)        & -112179(3)        &   -112000(280000) &-&-\\
$10^9   H_{K}''$    &4470000(2500)      &4470000(4100)      &4500000(200000)    &-&-\\
$10^9   h_{J}''$    &   45.3(14)        & 40.1(9)           &       42.3(17)    &-&-\\
$10^9   h_{JK}''$   &   15592(5)        & 15612(6)          &     15665(310)    &-&-\\
$10^9   h_{K}''$    &   1372119.6(24)   & 1372119.5(36)     &   1372000(18000)  &-&-\\
%\hline
%$\nu_{0}$                &  909572633(41)&  919120592(48)&  - &909571763(1110)&919120745(690)\\
%$\Delta A$               &   -35568.1(40)&   -22859.8(43)& -&-&-\\
%$\Delta B$               &    -5678.5(18)&    -5883.8(32)& -&-&-\\
%$\Delta C$               &    -3733.9(23)&    -4238.2(39)& -&-&-\\
%$10^3  \Delta D_{J}$     &       70.5(44)&          13(1)& -&-&-\\
%$10^3  \Delta D_{JK}$    &       1914(51)&        326(14)& -&-&-\\
%$10^3  \Delta D_{K}$     &     -62448(62)&      -9950(98)& -&-&-\\
%$10^3  \Delta d_{J}$     &      -19.6(40)&       12.7(27)& -&-&-\\
%$10^3  \Delta d_{K}$     &     15325(695)&      -838(383)& -&-&-\\
%$10^9  \Delta H_{J}$     &     6449(9970)&        -17(18)& -&-&-\\
%$10^9  \Delta H_{JK}$    &  -49136(12040)&      958(5400)& -&-&-\\
%$10^9  \Delta H_{KJ}$    &  265765(79629)&  270475(87700)& -&-&-\\
%$10^9  \Delta H_{K}$     &4298751(618191)&-1224662(642744)&-&-&-\\
%$10^9  \Delta h_{J}$     &   -24129(8455)&         11(11)& -&-&-\\
%$10^9  \Delta h_{JK}$    &    12464(2116)&    -1399(6167)& -&-&-\\
%$10^9  \Delta h_{K}$     &1761979(491145)& 113237(107326)& -&-&-\\
\hline
$\nu_{0}$           &  909572633(41)    &  919120592(48)    & -&909571763(1110) &919120745(690)\\
$A'$                & 246402.8(40)      & 259111.6(43)      & -&  246669(90)    &259090(21)\\
$B'$                & 33158.0(19)       & 32952.3(33)       & -&   33194(21)    &32918(16)\\
$C'$                & 30268.8(23)       & 29764.4(39)       & -&   30277(21)    &29837(14)\\
$10^3  D_{J}'$      &      145.9(44)    &        89.6(9)    & -&     188(17)    &99(12)\\
$10^3  D_{JK}'$     & 3204(52)          & 1617(14)          & -&   4886(450)    &1484(45)\\
$10^3  D_{K}'$      & -43024(63)        & 9474(98)          & -&-48237(1919)    &9383(33)\\
$10^3  d_{J}'$      &       -9.8(40)    & 23.1(27)          & -&      49(14)    &10(12)\\
$10^3  d_{K}'$      & 16350(700)        & 190(380)          & -& 11892(4497)    &-6175(959)\\
$10^9  H_{J}'$      & 6480(9970)        &         16(18)    & -&-&-\\
$10^9  H_{JK}'$     & -20000(12000)     & 30000(5400)       & -&-&-\\
$10^9  H_{KJ}'$     & 150000(80000)     & 158000(88000)     & -&-&-\\
$10^9  H_{K}'$      &   8770000(620000) & 3240000(640000)   & -&-&-\\
$10^9  h_{J}'$      & -24100(8500)      &         51(12)    & -&-&-\\
$10^9  h_{JK}'$     & 28100(2100)       & 14200(6200)       & -&-&-\\
$10^9  h_{K}'$      &3130000(490000)    &1490000(107000)    & -&-&-\\
\hline\hline
\end{tabular}
\end{center}
}
\end{table*}
\clearpage

%%%%%%%%%%%%%%%%%%%%%%%%%%%%
%%% Conclusion & Outlook %%%
%%%%%%%%%%%%%%%%%%%%%%%%%%%%
\section{Conclusion and Outlook}
Doppler-limited measurements covering the $2^1_04^3_0$ and $2^2_04^1_0$ rovibrational bands of formaldehyde between 30140\,cm$^{-1}$ and 30800\,cm$^{-1}$ were performed. The enhanced resolution compared to previous studies enables precise determinations of line positions and line strengths. In the Doppler-broadened spectra we find no indications for additional line broadening effects. Doppler-free measurements performed with small modifications to the experimental setup confirm previous studies of excited state lifetimes. The comparison of measured line positions to simulations of the spectrum using literature values for rotational constants shows significant deviations. Using genetic algorithms for the simulation of the rotational spectrum, a better agreement with measured data over a wide range of the spectrum is found. However some regions had to be excluded from the fit indicating perturbations of the rotational structure.

Using the detailed understanding about the rotational structure of the formaldehyde ultraviolet spectrum we have performed internal state diagnostics of guided cold formaldehyde beams \cite{Motsch2007}.

\section*{Acknowledgments}
We acknowledge assistance of A. Pérez-Escudero in the initial state of the experiment.
Support by EUROQUAM (Cavity-Mediated Molecular Cooling) and by the
Deutsche Forschungsgemeinschaft through the excellence cluster
"Munich Centre for Advanced Photonics" is acknowledged.

\clearpage

%%%%%%%%%%%%%%%%%%%%
%%% Bibliography %%%
%%%%%%%%%%%%%%%%%%%%

\end{document}